\documentclass[aps,twocolumn,showpacs,pra]{revtex4}
\usepackage{dcolumn}
\usepackage{graphicx}
\usepackage{amssymb}
\usepackage{epstopdf}
\usepackage[english]{babel}
\usepackage[utf8x]{inputenc}
\usepackage{wrapfig}
\usepackage{lipsum}
\usepackage{enumitem}

\begin{document}

\title{document}
\title{Unconventional electromagnetic properties of the graphene quantum dots%
}
\author{S.~E.~Shafraniuk}
\affiliation{Tegri LLC, Evanston, IL 60202}
\pacs{DOI: 10.1109}

\begin{abstract}
Quantum dots based on the graphene stripes show unconventional optical
properties in the THz frequency range. The graphene quantum dot (GQD) is
made of electrically gated stripe with zigzag edges. Inside the active
region (AR), which is enclosed between the source and drain electrodes,
there are two sharp energy ($\pm $)-levels, whose separation $2\Delta $ is
controlled with Stark effect by applying the lateral dc electric field. Such
the edge states determine the unique nature of elementary excitations,
chiral fermions, that are responsible for the non-linear optical responce
revealing a potential for many applications. They are, e.g., the frequency
multiplication and self-focusing of two dimensional solitons. Furthemore,
when injection of the non-equilibrium electrons causes an inverse population
of the levels localized in AR, the subsequent recombination of electrons and
holes leads to a coherent emission of the THz waves.
\end{abstract}

\maketitle

Unconventional optical properties of graphene in the THz range with
frequencies $f=0.5-100$~THz attract significant attention of many
researchers \cite%
{Gr_THz_A,Gr_THz_B,IBM-polaritons,TEbook,Rinzan,Kawano,Novoselov}. Interest
to the THz waves (T-rays) is motivated by variety of potential applications
in medicine, information technology, communication and security. One example
of the T-ray application is the remote sensing of chemical and biological
substances that requires powerful THz lasers and high-resolution spectral
analyzers. There are also suggestions of quatum dot THz detectors \cite%
{Rinzan,Kawano}, frequency multipliers \cite%
{S-A-Mikhailov,K-L-Ishikawa,E-Hendry} and self-focusing of two dimensional
solitons \cite{R-W-Boyd} in the electrically tunable metastructures. 
\begin{figure}[tbp]
\includegraphics[width=85 mm]{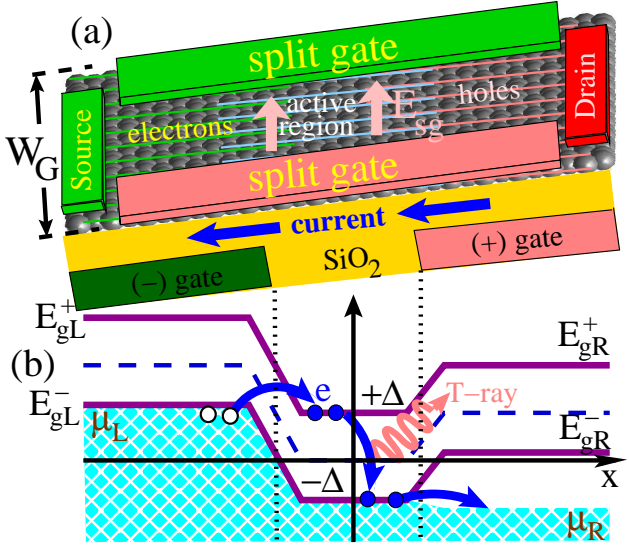} 
\caption{ {Color online. (a)~Geometry of the all-electrically controlled
quantum dot (GQD) made of the graphene stripe with zigzag edges. The split
gates form a transversal electric field $\mathbf{{E}_{\mathrm{sg}}}$, which
due to Stark effect splits the energy 0-level into the upper (with energy $%
+\Delta $) and lower [with energy $-\Delta $, see sharp peaks in the
electron density of states in Fig.~2d that originate from the ($\pm $%
)-levels], whose inverse population is created when a finite bias $%
V_{SD}\geq 2\Delta $ is applied between the source and drain electrodes
provided that the electron electrochemical potentials $\protect\mu _{L,R}$
in the left and right banks of the graphene stripe are set respectively as $%
\protect\mu _{L}=E_{\mathrm{gL}}^{-}$ and $\protect\mu _{R}\leq E_{\mathrm{gR%
}}^{-}-\Delta $ using the back ($\pm $)-gates. (b)~Energy diagram of GQD. }}
\label{Fig_1}
\end{figure}
Recently reported \cite{Rinzan,Kawano} carbon nanotube THz receivers and
spectral analyzers exploit the field-induced single-electron tunneling and
transitions between the quantized electron levels. Furthermore, unique
plasmonic characteristics of graphene allow building the tunable THz lasers 
\cite{Novoselov}. The surface plasmons (SP), whose spectrum changes versus
the applied gate voltage have been observed in several experiments \cite%
{Ju-2011,Eberlein,Kim-2011,J-Chen,Fei,Grigorenko,Yan,Wang,Polat,Novoselov-2012,Barnes} on graphene. 
\begin{figure}[tbp]
\includegraphics[width=85 mm]{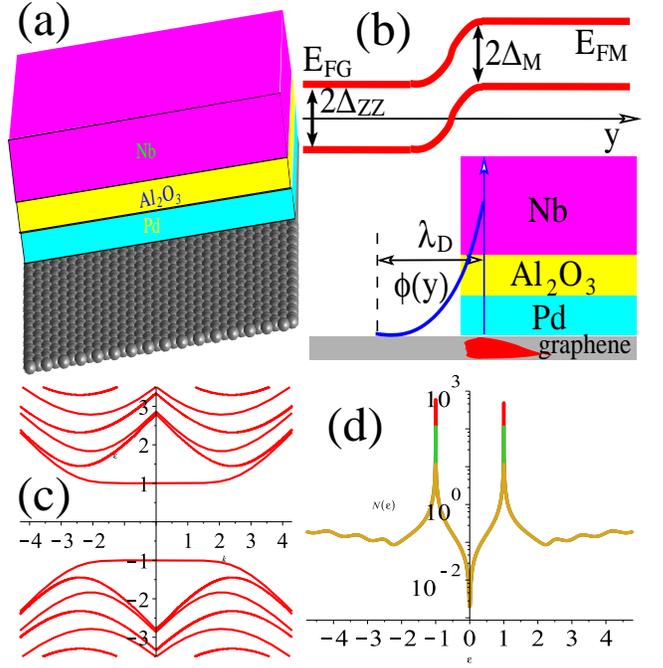} 
\caption{ {Color online. (a) One side of the Nb/Al$_{2}$O$_{3}$/Pd/G gate
which is deposited on the graphene stripe to form the GQD active region
shown in Fig.~1. (b)~The energy diagram at the graphene/gate boundary. (c)
Electron spectrum showing the subband structure of GQD made of the graphene
stripe enclosed between two timber-like split gate electrodes to form the
active region with effective zigzag edges shown in Fig.~\protect\ref{Fig_1}%
a. (d) The local electron density of states $D(\protect\varepsilon )$ inside
the GQD active region. The transversal dc electric field $\mathbf{E}_{%
\mathrm{sg}}$ due to the Stark effect splits the zero-energy peak into two
separate peaks forming an energy gap $2 \Delta $.}}
\label{Fig_2}
\end{figure}

One remarkable feature of graphene is that the carrier concentration, the
electrochemical potential $\mu $, and hence its conductivity $\sigma $ can
be appropriately tuned, e.g., by applying appropriate electric potentials to
the gate electrodes \cite{Ju-2011,Ozbay,Docherty,Das}. Hence, optical
properties of the 2D atomic monolayers are readily tunable in the terahertz
(THz) spectral region \cite{Ju-2011,Bonaccorso}, enabling their application
in the compact electrically controllable THz optical devices \cite%
{Tredicucci}. These opens new opportunities as compared to the noble metals
that are typically used in THz devices \cite{Ozbay}. Hence, the SP spectrum
in the 2D atomic monolayers is altered in situ, without any changes in the
device's design while optical and plasmonic characteristics of the 2D
materials are tunable in the terahertz (THz) spectral region \cite%
{Ju-2011,Bonaccorso}.

Currently, there are several concepts of the THz emitters made of the carbon
nanotube and graphene \cite{Ju-2011,Eberlein,Kim-2011,J-Chen}. A lot of
attention is paid to the solid state laser involving localized quantum
states arising in systems with reduced dimensionality, e.g., 2D (quantum
wells) or 1D (quantum dots). One example are the quantum cascade lasers
(QCL) based on layered semiconducting superlattices where series of quantum
wells with 2D electron spectrum are created. A "proof of concept" tunable
THz laser based on the gain modulation by graphene plasmons in an aperiodic
lattice and exploiting the unique properties of graphene plasmons was built
and tested in Ref.~ \cite{Novoselov}. However, despite their remarkable
performance, such the quantum cascade lasers (QCL) have serious setbacks.
Basic problem is that the energy dissipation caused by electron-phonon and electron-electron
collisions leads to considerable intrinsic Joule heating raising the internal local temperature far above the temperature of the external environment. Such the local Joule heating causes an adverse negative effect
on the QCL performance. To reduce the negative effect of the local heating,
one should decrease the bias current below some threshold value and also
cool down the QCL structure below ~200 K. The above measures complicate the
QCL design and limit the QCL system power and efficiency. Also there are
several concepts of the THz emitters made of the carbon nanotube and
graphene \cite{Ju-2011,Eberlein}. A "proof of concept" tunable THz laser
based on the gain modulation by graphene plasmons in an aperiodic lattice
and exploiting the unique properties of graphene plasmons was built and
tested in Ref.~ \cite{Novoselov}. 

A possible fundamental solution to the above issue of overheating represent 
systems with lowered dimensionality, e.g., with 1D or even 0D instead of 2D.
The electron bands in the 1D and 0D systems are much narrower than in the 2D
systems, which also means that the phase space where the electron-phonon
scattering occurs is reduced and most of the electron-phonon scattering
processes are eliminated. Hence, in the 1D and 0D systems, the intrinsic
energy dissipation due to suppression of the electron phonon scattering is
considerably lowered as well. This motivates the interest to electromagnetic
properties of low-dimensional  comprising quantum dots. Promising examples are the novel 2D atomic monolayers like graphene and its allotropes. The graphene stripes and carbon nanotubes represent 1D systems, whose dimensionality is reduced further to 0D by introducing an additional confinement by placing electrodes and local gates. Basically, 
quantum dot QD lasers are good candidates for the next generation high-speed
communication and already are more promising than quantum well lasers with
respect to important features like threshold current, temperature stability,
chirp, and feedback insensitivity \cite{Bimberg1,Bimberg2,Bimberg3}.
However, there is a need to understand what limits the performance and how
it can be improved. This requires a better understanding of the underlying
dynamics on a microscopic level. Below we study a graphene quantum dot
using microscopic approach for calculating the optical susceptibility.    

In this work we consider an all-electrically controlled 0D quantum dot based
on the graphene stripe with zigzag edges \cite{Shafr-Graph-Book} that
comprises a plasmonic THz microcavity. The motivation for this work is the
recent success in synthesis of graphene stripes with perfect zigzag edges 
\cite{Swiss-ZZ} where according to Refs. \cite%
{Fertig-1,Fertig-2,Shafr-5,Arabs} the topologically protected, sharp and
voltage-controlled the edge energy levels emerge. Below we will see, that
exploiting such the stable, voltage-controlled edge energy levels opens new
opportunities for designing the tunable THz devices. We utilize the unique intrinsic properties of graphene that allow for building various devices with novel remarkable properties. The study is focused on the ability of graphene quantum dot (GQD) to dynamically modulate round-trip modal gain values and shows potential to forming the laser emission. Such  the gated 2D monolayer material serves as a powerful tool to controlling the optical properties of GQD. The GQD device is instantly tunable and is all electrical in nature, with minimal electrical power demands.

The goal of this work is to compute the optical susceptibility of the graphene quantum dots (GQD) that describe their unconventional electromagnetic (EM) properties. The efforts are focused on the all-electrically controlled GQD fabricated using the graphene stripes with atomic zigzag edges. The knowledge of how the optical susceptibility of GQD depends on the frequency and electrochemical potential allows better understanding the physical mechanisms related to the electrically controlled absorption and emission of the electromagnetic field. Furthermore, the computation results allow finding, e.g., the conditions to the THz waves emission by the all-electrically controlled GQD. Furthermore our study also focuses on finding the non-linear electromagnetic response of GQD. We will see that the physical mechanism of such the non-linearity originates from the unconventional properties of chiral fermions in graphene stripes with atomic zigzag edges.

\section{The model}

Geometry of the proposed device is shown in Fig.~\ref{Fig_1}a where the
central part is the graphene stripe, whose properties are controlled by the
source drain and gate electrodes. The active region represents the graphene
quantum dot containing two sharp ($\pm $)-levels originating from the edge
states and spaced by 2$\Delta $. Since the edge states are topologically
protected \cite{Arabs}, the ($\pm $)-levels are very sharp and robust, even
if the edge roughness and impurities are present. The magnitude of the level
spacing 2$\Delta $ is controlled using the Stark effect by setting the
electric voltage between the split gates \cite{Shafr-5}. Besides, the energy
level positions $E_{n}$ in the left (L) and right (R) side sections of the
graphene stripe are controlled by applying electric potentials $V_{\mathrm{%
GL,R}}$ to the left and right side bottom gate electrodes respectively.
Furthermore, by applying a finite bias voltage between the source and drain
electrodes one injects the non-equilibrium electrons into the upper level
with energy $E_{+}=+\Delta $, thereby creating an inverse population of the
upper ($+$)-level in the active region. The electrons residing on the upper $\left( +\right) $ level then recombine to the lower $(-)$ level by emitting THz photons. Hence, the subsequent recombination of the electrons into the lower level with $E_{-}=-\Delta $ leads to a emission of the T-rays with frequency $f=2\Delta/h $. The interaction between light and material is controlled by the shape of the electromagnetic density-of-states (DOS) in the micro resonator \cite{Yablonovich,Pickering}. These mean that the magnitudes of lasing frequency $f$ and the amplification of resonant modes are set by applying the split-gate voltage $V_{\mathrm{SG}}$ across the active region of the laser and/or by the source drain and bottom gate voltages as shown in Fig.~\ref{Fig_1}. The latter mechanism is studied in details below in Sec.~V. We will see that the resonant frequency of THz emission depends not only on the split gate voltage $V_{\mathrm{SG}}$ as mentioned above but also it varies versus the electrochemical potential $\mu $, which is controlled by applying voltage to the gate electrodes. This enables flexible all-electric manipulations the lasing emission parameters \cite{Dowling,Engel,Li}.

Understanding the mechanisms determining the optical properties of GQD is
accomplished using solutions the Dirac equation complemented by appropriate
boundary conditions (see Refs. \cite%
{Fertig-1,Fertig-2,Shafr-5,Arabs,Shafr-Graph-Book,TEbook} for details). 

\section{Susceptibility of graphene}

Initially we consider a simplest case of the two-dimensional (2D) electron
gas in the atomic monolayer representing a suspended pristine graphene.
A general expression for the optical susceptibility of free carriers in
graphene that are chiral fermions (HF) takes the form%
\begin{equation}
\chi \left( \mathbf{{q},\omega }\right) =\frac{\left\vert d_{cv}\right\vert
^{2}}{L^{2}}\sum_{s,s^{\prime }=\pm 1}\int \frac{d^{2}\mathbf{k}}{\left(
2\pi \right) ^{2}}\frac{f\left( \epsilon _{s,\mathbf{{k}+{q}}}\right)
-f\left( \epsilon _{s^{\prime },\mathbf{k}}\right) }{\omega -\epsilon _{s,%
\mathbf{{k}+{q}}}+\epsilon _{s^{\prime },\mathbf{k}}+i\eta }  \label{kappa1}
\end{equation}%
where $d_{cv}$ is the electric dipole matrix element, whose indices $c,v$
are attributed to the conducting/valence bands, $L$ is the size of a
square-shaped 2D sample, $\eta $ is the damping parameter associated with
the electron energy dissipation during the inelastic collisions, $\mathbf{k}$
is the 2D electron momentum, $\mathbf{q}$ and $\omega $ are the respective
electron momentum and energy change, $f\left( \epsilon _{s,\mathbf{k}%
}\right) $ is the HF distribution function that depends on the HF excitation
energy $\epsilon _{s,\mathbf{q}}$, which for the pristine graphene conform
the continous dispesion law 
\begin{equation}
\epsilon _{s,\mathbf{q}}=sv_{F}\left\vert \mathbf{q}\right\vert
\label{E_cont}
\end{equation}%
where $s$ and $s^{\prime }=\pm 1$ are the HF branch indices, $v_{F}$ is the
Fermi velocity in graphene. The damping parameter $\eta $ in Eq. (\ref%
{kappa1}) actually plays the same role as parameter of the adiabatic
switch-on. For the HF spectrum (\ref{E_cont}), the density of states is \cite{Shafr-Graph-Book} %
\begin{equation}
D_{G}\left( \epsilon \right) =\frac{3\sqrt{3}a^{2}}{\pi v_{\mathrm{F}}^{2}}%
\left\vert \epsilon \right\vert \text{,}  \label{NE}
\end{equation}%
where $a$ is the lattice constant in graphene. 

The calculation details of the optical susceptibility are given in Appendix A. 
\begin{figure}[tbp]
\includegraphics[width=85 mm]{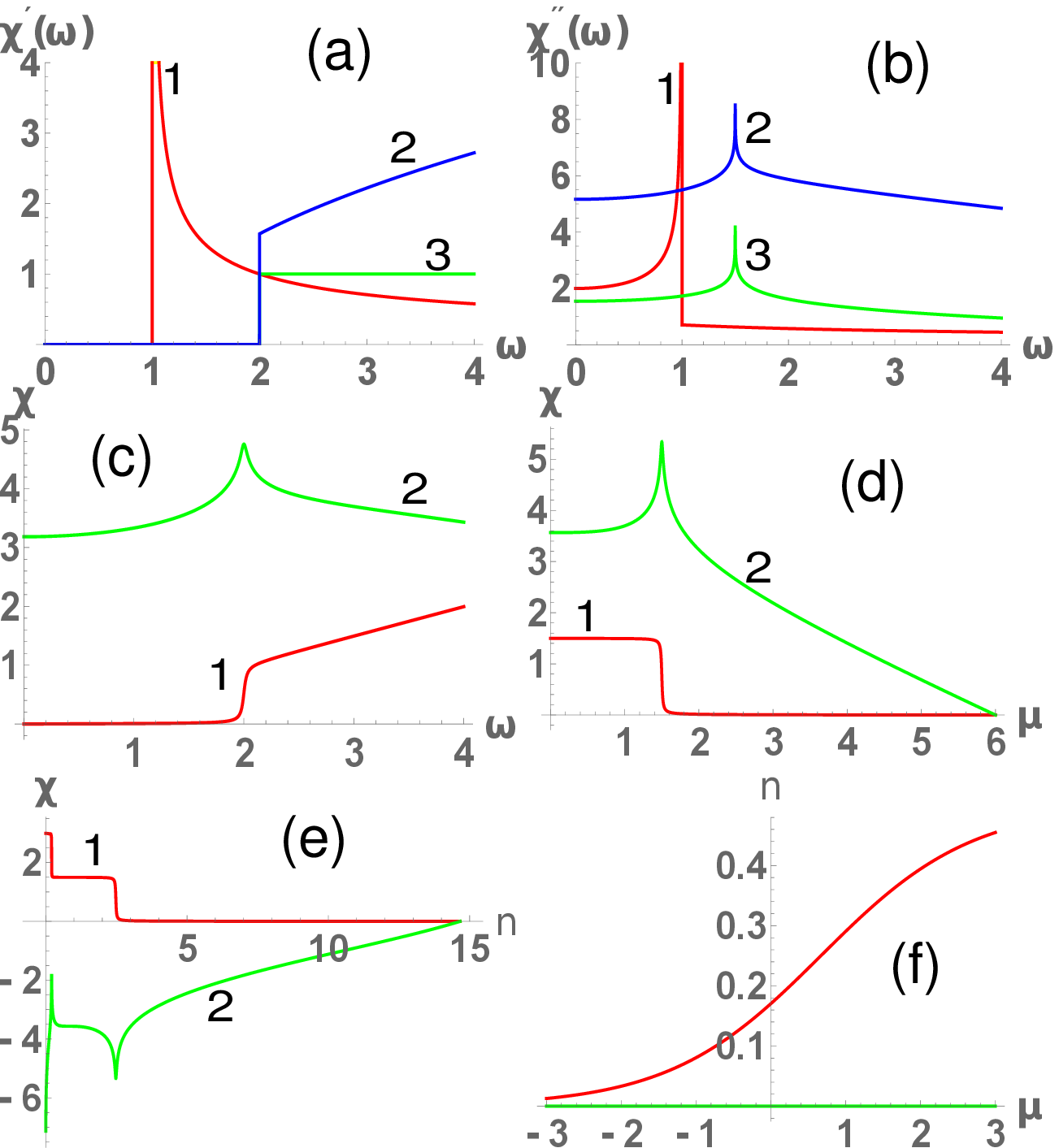} 
\caption{ {Color online. Optical susceptibility: (a) $\Re \protect\chi %
\left( \protect\omega \right) $ for $D=1..3$ (respective curve labels are 1,
2 and 3), $\protect\delta =1$, $\protect\mu =0.5$, and $\Lambda =36$. (b) $%
\Im \protect\chi \left( \protect\omega \right) $ for $D=1..3$ (respective
curve labels are 1, 2 and 3), $\protect\delta =1,$ $\protect\mu =0.5$, and $%
\Lambda =16$. (c) and (d) The real (curve 1) and imaginery (curve 2) parts
of the pristine graphene susceptibility $\protect\chi \left( \protect\omega %
\right) $ (see (c) for $\protect\mu =1$) and $\protect\chi \left( \protect%
\mu \right) $ (see (d) for $\protect\omega =3$) respectively versus the
frequency $\protect\omega $ and electrochemical potential $\protect\mu $ for 
$\eta =0.02,$ $\Lambda =6$. (e) The real (curve 1) and imaginary (curve 2) 
$\protect\chi \left( n\right) $ for the pristine graphene for $\eta =0.2$, 
$\Lambda =6$, $\protect\mu =1$. (f) The HF excitations concentration $n(%
\protect\mu )$ versus the electrochemical potential $\protect\mu $ for
pristine graphene.}}
\label{Fig_3}
\end{figure}
Results of the calculation are illustrated in Figs.~\ref{Fig_3} where we
show the real $\chi ^{\prime }=\Re \chi $ and imaginary $\chi ^{\prime
\prime }=\Im \chi $ parts of the optical susceptibility versus the frequency 
$\omega $ and electrochemical potential $\mu $. 
As an illustaration, we also show results for a conventional semiconductor with dimensionalities $D = 1, 2, 3$ (see Fig.~\ref{Fig_3}a,b). Parameters of the calculations, whose details are given in Appendix A, are indicated in the caption. Respective results for $\Re \protect\chi %
\left( \protect\omega \right) $ (curve~1) and $\Im \protect\chi %
\left( \protect\omega \right) $ (curve~2) of the pristine graphene are shown in  Fig.~\ref{Fig_3}c while Fig.~\ref{Fig_3}d illustrates results for $\Re \protect\chi %
\left( \protect\mu \right) $ (curve~1) and $\Im \protect\chi %
\left( \protect\mu \right) $ (curve~2). Figure Fig.~\ref{Fig_3}e shows the results for $\protect\chi $ versus the charge carrier concentration $n$, as indicated in the caption. The last Fig.~\ref{Fig_3}f shows dependence of the HF charge carrier concentration $n$ versus the electrochemical potential $\protect\mu $ for pristine graphene.

At finite temperatures and in the non-equilibrium conditions,
the respective calculations of $\chi \left( \omega ,\mu \right) $ are
conducted numerically. Numeric solutions are also useful when considering
graphene-based structures with more complicated excitation spectrum. Below
we study the optical properties of the graphene stripe with zigzag edges and
non-linear optical properties of GQD when a strong ac field is present.

\section{ZZ stripe of graphene}

For graphene stripes, whose width $W$ is finite, one should also consider
quantization of the HF excitations in the lateral direction. Below we
compute optical susceptibility of the graphene stripe with atomic zigzag
edges. We use units with $\hbar =1$ and $k_{\mathrm{B}}=1$
if not stated otherwise.  The edge states \cite%
{Fertig-1,Fertig-2,Shafr-5,Arabs,Shafr-Graph-Book,TEbook} emerging in the
graphene stripe change the HF excitation spectrum significantly. This
happens because the chirality of excitations in graphene imposes additional
constrains on reflection processes due to conservation of two pseudospins. A transverse 
d.c. electric field $\mathbf{E}=\{0,E_{y},0\}$ is applied perpendicular to the graphene stripe axis utilizing the split gate electrodes shown in Fig. \ref{Fig_1}a. When $E_{y}=0$, the
electron reflections at the atomic zigzag edges \cite%
{Fertig-1,Fertig-2,Shafr-5} cause a crisp narrow energy level to arise at
the energy $\epsilon =0$. When the transverse electric field is finite, $%
E_{y}\neq 0$, such the zero-energy level experiences Stark splitting, whose
magnitude is $2\Delta =2eE_{y}W$ (see Refs. \cite%
{Shafr-5,Arabs,Shafr-Graph-Book,TEbook}). Then, the Stark splitting of the
sharp singularities at energies $\epsilon =\pm \Delta $ \ emerging in the
electron density of states (see Fig.~\ref{Fig_2}d) is controlled by the
electric field $E_{y}\neq 0$. Below we will see that such the electrically
controlled HF spectral singularities are responsible for the remarkable
optical properties of the graphene quantum dots. The effect is described in
terms of the susceptibility $\chi (\omega ,\mu )$, which we compute below. An additional control of $\chi (\omega ,\mu )$ is introduced with the top (or bottom) local gate electrodes (see Fig. \ref{Fig_1}a). Thus, the respective local gate voltages control the both, the Stark splitting $\Delta $ along with the HF electrochemical potential $\mu $. Below we will see that the shape of $\chi
(\omega ,\mu )$ depends on both, $\mu $ as well as on $\Delta $.

There is no simple analytical expression for the HF excitation energy in the
graphene stripe with atomic zigzag edges~\cite%
{Fertig-1,Fertig-2,Shafr-5,Arabs,Shafr-Graph-Book,TEbook}. The HF dispersion
law~\cite{Fertig-1,Fertig-2,Shafr-5,Arabs,Shafr-Graph-Book,TEbook} is given
by the two equations 
\begin{equation}
\epsilon _{k}=\mu \pm \sqrt{\Delta ^{2}+k_{y}^{2}v_{\mathrm{F}}^{2}+k_{x}^{2}v_{%
\mathrm{F}}^{2}}  \label{Egr}
\end{equation}%
\begin{equation}
k_{y}=\frac{k_{x}}{\tan \left( Wk_{x}\right) }  \label{kt}
\end{equation}%
that describe the constrain due to the pseudospin conservation during
elastic reflections of the HF excitations on atomic edges. The electron energy in graphene stripe (\ref{Egr}) depends on the two components of the electron (HF) momentum, $k_{y}$ (longitudinal) and $k_{y}$ (transversal), which are related by Eq. (\ref{kt}). In an infinitely long stripe by width $W$, $k_{y}$ is continuous while $k_{x}$ discrete, because the transversal momentum is quantized. In the geometry on Fig. \ref{Fig_1}a, the active region length $L_{a}$ is $L_{a}>>W$, therefore the quantization along the $y$-axis  is negligible. We will see that such the constrain allows reducing the
dimensionality of the system from 2D as in pristine graphene to 1D for the stripe. In Eqs. (\ref{Egr}), (\ref{kt}), $\epsilon _{k}$ is the HF energy variable, $\mu $ is the electrochemical potential, $\Delta =eE_{y}W$ is the Stark splitting of the zero-energy level posing as the "energy gap" in Eq. (\ref{Egr}). In such the case, the splitting energy also depends on the gate efficiency 
$\alpha $. Magnitude of $\mu $ is controlled by the bottom (or top) gate electrodes, while $\Delta $ is
controlled by the split gate electrodes depicted in Fig. \ref{Fig_1}a as
illustrated by energy diagram in Fig. \ref{Fig_1}b. The HF dispersion law is
computed by solving Eqs. (\ref{Egr}), (\ref{kt}). The HF excitation energy $%
\epsilon (k_{y})$ in the graphene stripe with zigzag edges is shown in Fig.~%
\ref{Fig_2}c versus the longitudinal momentum $k_{y}$. Technically, the
electron density of states shown in Fig. ~\ref{Fig_2}d is computed as%
\begin{equation}
D_{\mathrm{ZZ}}\left( \epsilon \right) =\left\vert \frac{dk_{y}\left(
k_{x}\right) }{d\epsilon _{k}}\right\vert =\left\vert \frac{dk_{y}\left(
k_{x}\right) }{dk_{x}}/\frac{d\epsilon _{k}}{dk_{x}}\right\vert \text{,}
\label{DOS_ZZ}
\end{equation}%
which gives an analytical expression%
\begin{equation}
D_{\mathrm{ZZ}}\left( \epsilon \right) =\left\vert \frac{\epsilon _{k}}{k_{x}%
}\frac{\tan k_{x}W-k_{x}W\left( \tan ^{2}k_{x}W+1\right) }{\tan ^{2}\left(
k_{x}W\right) }\right\vert \text{,}  \label{DOS}
\end{equation}%
where according Eqs. (\ref{Egr}), (\ref{kt}), $k_{x}$ depends on the energy variable $\epsilon $. To compute $\chi (\omega ,\mu )$ for the quasi-1D graphene stripe with atomic zigzag
edges, we again use the general expression Eq. (\ref{kappa1}). The
calculation is simplified for the direct interband transitions ($\mathbf{q}%
=0 $). 
Then we get 
\begin{eqnarray}
\chi \left( \omega \right) &=&\frac{\left\vert d_{cv}\right\vert ^{2}}{2\pi
a^{2}}\int \frac{\left[ f\left( \epsilon _{k}-\mu \right) +f\left( \epsilon
_{k}+\mu \right) -1\right] }{\omega \pm 2\epsilon _{k}+i\eta }dk_{y}dk_{x} 
\nonumber \\
&=&\frac{\left\vert d_{cv}\right\vert ^{2}}{2\pi a^{2}}\int_{k_{\mathrm{max}%
}}^{k_{\mathrm{min}}}D_{\mathrm{ZZ}}\left( k_{x}\right) \left[ 1-f\left(
\epsilon _{k}-\mu \right) -f\left( \epsilon _{k}+\mu \right) \right] 
\nonumber \\
&&\times \left( \frac{1}{\omega -2\epsilon _{k}+i\eta }-\frac{1}{\omega
+2\epsilon _{k}+i\eta }\right) dk_{x}\text{,}
\label{kappa6}
\end{eqnarray}
where we use%
\begin{eqnarray}
f\left( \epsilon _{s,k}\right) -f\left( \epsilon _{s^{\prime },k}\right)
\nonumber \\
=f\left( -\mu +\epsilon _{k}\right) -f\left( -\mu -\epsilon _{k}\right) =
\nonumber \\
\pm \left[ f\left( \epsilon _{k}-\mu \right) +f\left( \epsilon _{k}+\mu \right)
-1\right]
\end{eqnarray}
and
\begin{equation}
dk_{y}=D_{\mathrm{ZZ}}(k_{x})dk_{x}\text{.}
\label{dkt}
\end{equation}%
When integrating (\ref{kappa6}) in infinite limits, the respective integral
diverges. Therefore we introduce cutoff by setting the lower $k_{\mathrm{min}}$ and upper $k_{\mathrm{max}}$ limits of integration in
Eq. (\ref{kappa6}), which respectively are found as solutions of the
equations 
\begin{equation}
k_{\mathrm{min}}^{2}=\mu ^{2}/v_{\mathrm{F}}^{2}-\left[ k_{y}\left( k_{%
\mathrm{min}}\right) \right] ^{2}-\left( \Delta /v_{\mathrm{F}}\right) ^{2}
\label{tmn}
\end{equation}%
and 
\begin{equation}
k_{\mathrm{max}}^{2}=\Lambda ^{2}/v_{\mathrm{F}}^{2}-\left[ k_{y}\left( k_{%
\mathrm{max}}\right) \right] ^{2}-\left( \Delta /v_{\mathrm{F}}\right) ^{2}%
\text{.}  \label{tmx}
\end{equation}%
Above we have used that the respective change of the HF excitation energy is 
$-\epsilon _{s,k}+\epsilon _{s^{\prime },k}=\pm 2\epsilon _{k}$, where $%
\epsilon _{\kappa }$ is defined by Eq. (\ref{Egr}). The above Eqs. (\ref{kt}%
), (\ref{tmn}) and (\ref{tmx}) serve as the closed system of transcendental
equations allowing to finding $k_{\mathrm{min}}$ and $k_{\mathrm{max}}$.
From Eq. (\ref{DOS_ZZ}) one can see that in contrast to pristine graphene,
whose the HF density of states (\ref{NE}) is a smooth function of the energy
variable $E$, the respective density of states $D_{\mathrm{ZZ}}(\epsilon )$
given by Eq. \ (\ref{DOS_ZZ}) for the graphene stripe with zigzag (ZZ) edges
shows dramatically different behavior, as illustrated in Fig.~\ref{Fig_2}d.
Namely, owing to appearance of ZZ edge states in the graphene stripe, the
respective singularities in the density of states (\ref{DOS_ZZ}) arise when $%
\tan ^{2}\left( k_{x}W\right) =0$ in denominator of Eq.~(\ref{DOS}) provided 
\begin{equation}
k_{x}=\frac{\pi }{W}m
\end{equation}%
where $m$ is integer. One can notice the mentioned sharp singularities at
energies $\epsilon =\pm \Delta $ (in units of Stark splitting $\Delta $) in
the plot of DOS on Fig. ~\ref{Fig_2}d., while the singularities are smoothed
at energies, $\epsilon >\Delta $ and $\epsilon <-\Delta $. The number of
excitations in the ZZ graphene stripe is computed as 
\begin{equation}
N=2\sum_{\mathbf{k}}f_{k}\rightarrow \frac{12\sqrt{3}a^{2}}{\pi v_{F}^{2}}%
\int_{k_{\mathrm{max}}}^{k_{\mathrm{min}}}D_{\mathrm{ZZ}}(k_{x})f\left(
\epsilon _{k}-\mu \right) dk_{x}\text{,}  \label{Npart}
\end{equation}%
where the lower and upper integration limits are again determined by
solution of Eqs. (\ref{kt}), (\ref{tmn}) and (\ref{tmx}).

The relevant energy scale in the above formulas (\ref{kappa6})-(\ref{Npart})
is determined by the atomic edge geometry and by the graphene stripe width $W$. Other energy scales in Eqs. (\ref{kappa6})-(\ref{Npart}) are related to $\omega $,  $\Delta $, $\mu $ and $\eta $. Typical magnitudes of interest here are $\omega \sim  2\pi f$, where for the electromagnetic field frequency $f=1~\mathrm{THz}$, the respective photon energy is $hf=4$ meV. Then,  $\mu \sim 2\Delta \sim hf=4$~meV, and we also use $\eta \sim  0.1\Delta $. Essentially, the last parameter, $\eta $, which also determines the width of quantized levels localized in the active region depends on coupling of GQD to the substrate and also by the inelastic  collisions, which also depend on the temperature and GQD geometry. 

\section{Electromagnetic emission from GQD}

A fundamental problem when designing the laser for the frequency region
0.5-100~THz is that the THz photon energy $E_{\nu }^{\mathrm{T}}$ is
relatively low, $E_{\nu }^{\mathrm{T}}\approx $ 4 meV - 0.4~eV, as compared
to a visible light photon for which $E_{\nu }^{\mathrm{L}}\approx $1.8 -
3~eV. Therefore, to ensure a monochromatic THz emission, the width $\eta $ of
quantized levels localized in the active region is required to be much
narrower than $\Delta $. This problem is solved by a proper designing the
active region, which is the key element of any solid state laser. Parameters
of the active region must satisfy to a number of requirements, which have to
be observed in order to get the T-ray emission out of it. In conventional
visible light lasers, $\eta $ is typically much smaller than the level spacing between the e/h-levels, i.e., $\eta
<<\Delta $. Then, the photon energy $E_{\nu }^{\mathrm{L}}$ is precisely
equal to the level spacing energy $\Delta $, i.e., $E_{\nu }^{\mathrm{L}%
}\equiv \Delta $ while the emitted light beam is fairly monochromatic and
coherent. The situation is different in the THz lasers where at the bottom
part of the THz domain the condition $\eta <<\Delta $ might fail if the level broadening $\eta $ exceeds $\Delta $, which is relatively small, $\Delta \approx 4$ meV. In the latter case, the spectrum of photon emission acquires the finite width (line broadening) while the photon energy distribution becomes dependent also on the width $\eta $ of the electron energy level $E_{\nu }$. Then the finite $\eta $ causes an extra decoherence and broadening the THz laser emission spectrum. In a worst case scenario one can even get $\eta \geq \Delta $, which causes complete violating of the condition $E_{\nu }^{\mathrm{T}}\equiv \Delta $. The latter example illustrates why creating the THz lasers is so difficult. Other sources the T-beam decoherence and the line broadening consist of the temperature fluctuations and noises which also strongly impact the THz device performance. Using of graphene suggests several possible solutions of the mentioned decoherence problem. (\textit{i})~One is able to form very narrow electron energy levels in the active region of the THz laser where $\eta <<\Delta $. Very sharp and narrow e/h-levels are obtained inside a narrow stripe of graphene with zigzag (ZZ) edges polarized by a transverse electric field as suggested in Refs. \cite%
{Fertig-1,Fertig-2,Shafr-5,Arabs,Shafr-Graph-Book,TEbook}. (\textit{ii})~One
can significantly reduce impact of phonons which essentially contribute
into the level width $\eta $. It is accomplished when using of a
narrow stripe having a definite orientation in respect to crystallographic
directions of the graphene lattice. The lattice symmetry and the atomic edges impose additional selection
rules on the scattering probability involving just phonons with certain
polarization and wave vector on the one hand and electrons with one-dimensional dispersion law on the other hand. Such the restriction rules out many inelastic scattering processes as irrelevant. Furthermore, due to
narrow width $\delta $ of the electron bands in GQD, the high-energy phonons with frequency $f_{ph} > \delta/h  $ do not participate in the electron-phonon scattering as well. The other mechanism involving the electron-electron collisions is less significant for graphene in the THz frequency range. The above results in very low energy dissipation in the graphene stripes with zigzag edges. (\textit{iii})~The levels originate from the topologically protected edge states \cite{Arabs}, hence they are robust in
respect to the lattice imperfections and thermal excitations.

Here we consider the limit of high density of the charge carriers, whose
energy recombination time $\tau_{\rm HF}$ is much shorter than the change in polarization of
the electromagnetic wave. The polarization relaxes very fast, and is
governed by the carrier--carrier and carrier--phonon scattering causing the
relaxation to its quasi-equilibrium value, which is determined by the
momentary magnitudes of the field and the carrier density. This ensures the
simplest \emph{quasi-equilibrium} conditions of a stationary excitation when
the carriers are in equilibrium with themselves while the graphene stripe is
out of equilibrium. In the quasi-equilibrium approximation, the field
intensity is a slow functions of time. We disregard all the effects causing
deviations from the quasi-equilibrium assumption, such as spectral, spatial,
or kinetic hole burning. This enables using the electron--hole--pair rate
equation complemented with laser specific terms. The \emph{rate equation}
for generating of $N$ photons takes the form 
\begin{equation}
\dot{N}=r_{p}-r_{st}-r_{sp}-r_{nr}\text{,}  \label{rate_eq}
\end{equation}%
where $r_{p}$ is the pumping rate, $r_{st}$ is the stimulated emission rate, 
$r_{sp}$ is the spontaneous emission rate, $r_{nr}$ is the non-radiative
transitions rate. The \emph{pump rate} due to injection current density $j$
is $r_{p}=j\eta_{QE} /\left( eW\right) $ where $\eta_{QE} $ is the quantum efficiency, 
$W$ is the transverse dimensions of the laser$^{\prime }$s active region
(i.e., the stripe width). The \emph{stimulated emission} \emph{loss} rate is 
$r_{st}=-\chi ^{\prime \prime }\left( \omega \right) \mathcal{E}%
_{0}^{2}/\left( 2\hbar \right) $ where $\omega =2\pi f$. The
quasi-equilibrium susceptibility $\chi \left( \omega \right) $ contains the
factor $1-f_{e,k}-f_{h,k}=\left( 1-f_{e,k}\right) \left( 1-f_{h,k}\right)
-f_{e,k}f_{h,k}$ that is conveniently separated in the two terms as $\chi
^{\prime \prime }\left( \omega \right) =\chi _{a}^{\prime \prime }\left(
\omega \right) -\chi _{e}^{\prime \prime }\left( \omega \right) $. Here $%
f_{e(h),k}$ is the distribution function of electron(hole)-like HF
excitations and the term $\chi _{e}^{\prime \prime }\left( \omega \right)
\propto f_{e,k}f_{h,k}$ describes the emission while the other term $\chi _{a}^{\prime \prime }\left( \omega \right)$ describes the absorption, $\chi _{a}^{\prime \prime }\left( \omega \right)
\propto \left( 1-f_{e,k}\right) \left( 1-f_{h,k}\right) $. The imaginary
part of susceptibility $\chi _{e}^{\prime \prime }\left( \omega \right) $ is
related to the gain $g\left( \omega \right) $ as 
\begin{equation}
g_{e}\left( \omega \right) =-\frac{4\pi }{n_{b}c}\omega \chi _{e}^{\prime
\prime }\left( \omega \right) \text{,}
\end{equation}%
where $g_{e}\left( \omega \right) $ is the probability per unit length to
emitting a photon and the background refractive index is $n_{b}\simeq \sqrt{%
\epsilon _{0}}$. Thus $-g_{e}c/n_{b}$ is the emission probability of a
photon per unit time. The \emph{spontaneous emission } rate into the
continuum of all photon modes with the frequency $\omega =\omega _{q,\lambda
}$ where $q,\lambda $ are the photon wave vector and polarization is 
\begin{equation}
r_{sp}=\frac{4}{\pi \epsilon _{0}}\int dqq^{2}\omega _{q}\chi _{e}^{\prime
\prime }\left( \omega _{q}\right) 
\end{equation}%
or 
\begin{equation}
r_{sp}=\frac{1}{2\pi \epsilon _{0}}\int_{0}^{\infty }d\omega \left( \frac{%
2\omega n_{b}}{c}\right) ^{3}\chi _{e}^{\prime \prime }\left( \omega \right)
.
\end{equation}%
The dependence $\sim \omega ^{3}$ indicates that the spontaneous emission
losses dominate at higher laser frequencies. The \emph{non-radiative}
emission rate is computed as $r_{nr}=N/\tau +CN^{3}$ where the 1-st term
corresponds to the multi-photon emission involving deep trap levels while
the 2-nd term might contain a significant contribution from the Auger
processes in the THz lasers.

The T-ray laser emission is described in terms of the semi-classical
electric field equation for spatial eigenmodes 
\begin{equation}
\left[ 1+4\pi \chi \left( N\right) /\epsilon _{0}\right] \ddot{\mathcal{E}}%
_{n}+\left( \kappa c/n_{b}\right) \dot{\mathcal{E}}_{n}+\omega _{n}^{2}%
\mathcal{E}_{n}=0  \label{field_eq}
\end{equation}%
where where $\omega _{n}$ is the eigenfrequency of the $n$-th resonator mode
and we have introduced the cavity loss rate as $\kappa c/n_{b}=4\pi \sigma
/\epsilon _{0}$ where $\sigma $ is the electric conductivity. For Eq.~(\ref%
{field_eq}), there are two regimes of the \emph{steady state} solutions: (%
\textit{i})~When the gain $g\left( \omega =0\right) $ is less than the
cavity losses, the laser field vanishes, i.e., for $\kappa >g\left(
N_{0},\omega _{m}\right) $ one gets the magnitude of time - averaged (i.e.,
at $\omega =0$) electric field $\mathcal{E}_{0}=0$ and $r_{p}=N_{0}/\tau $,
where $N_{0}$ is the time - averaged number of photons. (\textit{ii})~If the
gain becomes equal or exceeds the cavity losses provided $\kappa =g\left(
N_{0},\omega _{m}\right) $, one gets the finite magnitude of laser field,
i.e., $\mathcal{E}_{0}\neq 0$. Namely, 
\begin{equation}
\mathcal{E}_{0}^{2}=\frac{2\hbar }{\chi ^{\prime \prime }\left( N_{0}\right) 
}\left( \frac{N_{0}}{\tau }-r_{p}\right)
\end{equation}
and 
\begin{equation}
\omega _{m}^{2}=\frac{\omega _{n}^{2}}{1+4\pi \chi ^{\prime }\left(
N_{0}\right) /\epsilon _{0}}\text{,}
\end{equation}%
where $\omega _{m}$ is the lasing frequency and $\chi ^{\prime }$ and $\chi
^{\prime \prime }$ are the real and imaginary parts of the optical
susceptibility in the active region, $\epsilon _{0}$ is the background
dielectric constant. The above formula for $\omega _{m}^{2}$ suggests that
the pulling of the laser mode is caused by the refractive index changes due
to the increased carrier density that is controlled by the gate voltages.

\section{Steady state susceptibility of GQD}

\begin{figure}[tbp]
\includegraphics[width=85 mm]{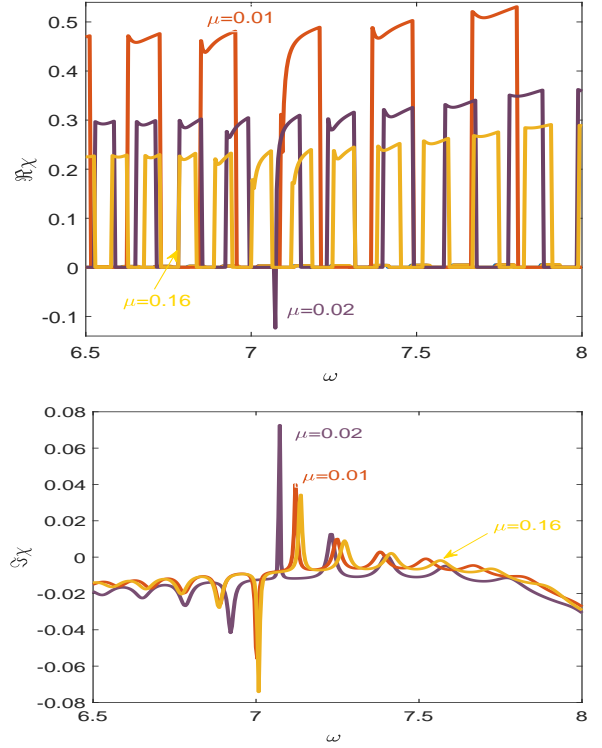} 
\caption{ {Color online. The steady state optical susceptibility $\Re 
\protect\chi \left( \protect\omega \right) $ (top panel) and $\Im \protect%
\chi \left( \protect\omega \right) $ (bottom panel) as function of frequency 
$\protect\omega = 2\protect\pi f$ in GQD computed for the stripe width $W =
2 h v_{\mathrm{F}}/ \Delta $, temperature $%
T=1.4$ (in units of $\Delta /k_{\mathrm{B}}$, $k_{\mathrm{B}}$ is Boltzmann
constant), the ineleastic collision rate $\eta =0.15$ and the level
spacing (Stark splitting) $2\Delta =2$ (all in units of $\Delta $). Here $%
\protect\omega = 2\protect\pi f$, $f$ is the frequency and $\protect\mu $ is
the electron electrochemical potential. The respective values of $\protect%
\mu $ are shown in figure. }}
\label{Fig_4_2nd}
\end{figure}

\begin{figure}[tbp]
\includegraphics[width=85 mm]{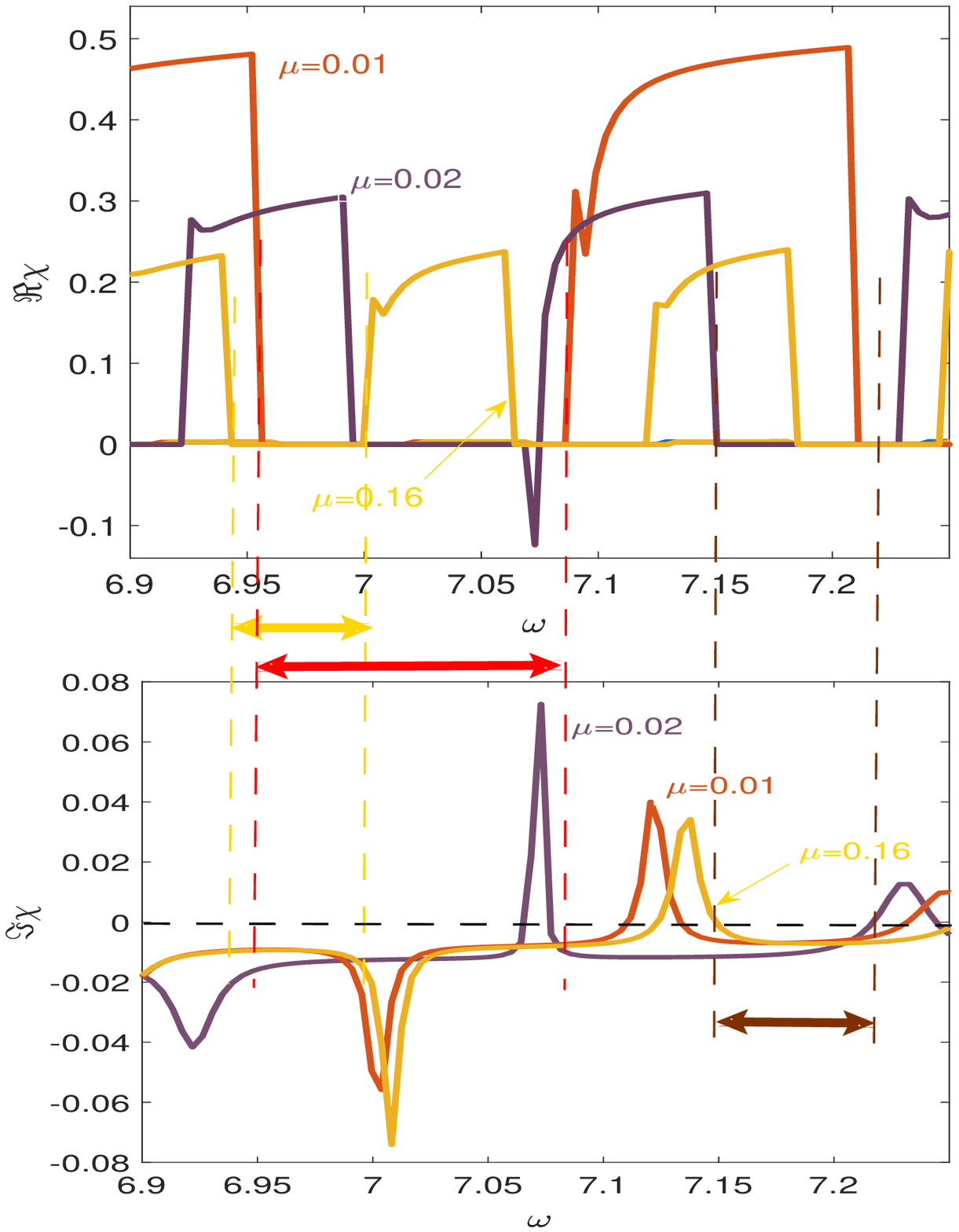} 
\caption{ {Color online. Blowup $\Re \protect\chi \left( \protect\omega %
\right) $ (top panel) and $\Im \protect\chi \left( \protect\omega \right) $
(bottom panel) shown in the former Fig.~\protect\ref{Fig_4_2nd} in the narrower
frequency interval. The instability regions corresponding to the conditions $%
\Re \protect\chi \left( \protect\omega \right) \vert_{\protect\omega =
\Omega_{\mathrm{p}}} = 0$ provided $\Im \protect\chi \left( \protect\omega %
\right) < 0 $ are marked by arrows (the arrow colors correspond to the curve collors). They correspond to a set of resonant frequencies $\Omega_{\mathrm{p}}$ determined by the mentioned conditions. }}
\label{Fig_5_2nd}
\end{figure}

Below we consider the effect of the electrochemical potential $\mu $ on the optical susceptibility of the graphene quantum dot (GQD), which is controlled by applying electric potentials to the back (or top) gate electrodes. The calculation results for the steady state optical susceptibility $\chi \left(\omega \right) $ are presented in Figs.~\ref{Fig_4_2nd}, \ref{Fig_5_2nd}, \ref{Fig_7} and \ref{Fig_8}.

The real $\Re \chi \left( \omega \right) $ and imaginary $\Im \chi \left( \omega \right) $ parts are shown in Fig.~\ref{Fig_4_2nd} as functions of frequency $\omega =2\pi f$ for different values of the electrochemical potential $\mu $, whose respective values are indicated in the
figures. One can see that the frequency dependence of $\chi \left( \omega
\right) $ dramatically changes as $\mu $
varies. Physically, this reflects the drastic change of the GQD optical
properties since the magnitude of $\mu $ determines the quantization
conditions at the graphene stripe edges. Remarkably, as $\mu $ changes, the
signs and magnitudes of the real and imaginary parts of $\chi \left( \omega
\right) $ alter. In Fig.~\ref{Fig_5_2nd}, we detalize the instability regions
in the narrower frequency intervals. One can see that in certain frequency
intervals the real part vanishes, $\Re \chi \left( \omega \right) = 0$, while $\Im \chi \left( \omega \right) <0$ remains negative. An important conclusion drawn from Figs.~\ref{Fig_4_2nd} and \ref{Fig_5_2nd} is that there is a set of resonant frequencies $\Omega _{\mathrm{p}}$ determined by the
condition $\Re \chi \left( \omega \right) |_{\omega =\Omega _{\mathrm{p}}}=0$
provided $\Im \chi \left( \omega \right) <0$. Remarkably, the $\Omega _{%
\mathrm{p}}$ magnitude depends on $\mu $ and $\Delta $, so in experments it
can be controlled by applying appropriate electric potentials to the gate
electrodes. Hence, based on the data presented in Figs.~\ref{Fig_4_2nd} and \ref%
{Fig_5_2nd}, one concludes that lasing conditions are fulfilled at $6.93<\omega
<7.03$, $6.96<\omega <7.09$ and $7.14<\omega <7.22$ (in units of $\Delta $). However, when $\Im
\chi \left( \omega \right) $ becomes positive (see peaks of $\Im \chi \left( \omega \right) $ in Fig.~\ref{Fig_5_2nd}), the lasing condition fails.

At the first sight it seems there is a problem with generating the coherent
THz radiation with the resonant frequency $\Omega_p \sim 6\Delta/hbar $ when $%
\Omega_p < 3$~THz, which requires the minimum level spacing $\Delta < 3 
\text{ THz}/6 = 0.5$~THz. Such the narrow level spacing $\sim 2$~meV
corresponds to $k_{\mathrm{B}}T \sim 20$~K, which is far below the room
temperature $T_{\mathrm{room}} \sim 300$~K. Deceptively, it seems that at $%
T=T_{\mathrm{room}} \sim 300$~K, the large temperature broadening $\eta \sim
30$~meV smears the spectral singularities at $\epsilon = \pm \Delta$ out
because it largely exceed the level spacing $2\Delta = 4$~meV, thereby
making the levels non distinguishable. However, according to detailed
calculations (see e.g., Ref.~\cite{Serhii-Spring}) for the narrow levels in
the graphene stripes with zigzag edges, the actual level broadening due to
the inelastic scattering is two orders of magnitude lower
than for conventional electrons with continuous dispersion law. The physical
reason is that the localized HF excitations interact with phonons very
weakly, because the respective phase space is confined. Therefore, the
actual broadening of the ($\pm $)~levels is very low and is below 3 meV even
at $T=T_{\mathrm{room}} \sim 300$~K. Thus, for narrower separation down to $%
2\Delta \sim 3$~meV in GQD that corresponds to $f \sim 0.7$~THz, the levels
remain well-defined even at room temperature.

\begin{figure}[tbp]
\includegraphics[width=85 mm]{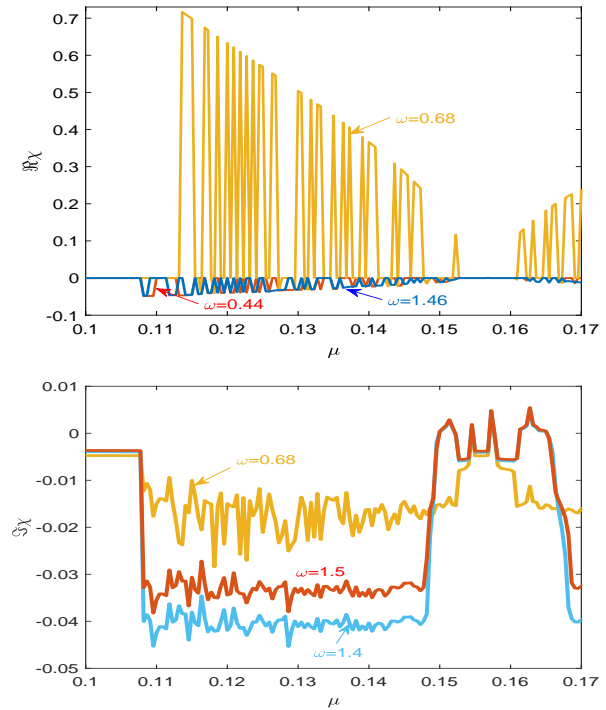} 
\caption{ {Color online. The steady state susceptibility $\Re \protect\chi %
\left( \protect\mu \right) $ (top panel) and $\Im \protect\chi \left(\protect\mu \right) $ (bottom panel) as function of electrochemical potential $\protect\mu $. The other parameters of GQD are the same as in Fig.~\protect\ref{Fig_4_2nd}. The respective values of $\protect\omega $ are shown in figure. Physically, the sharp switching occur when $\mu $ hits the quantized level positions, which are dense for the relatively broad stripe.}}
\label{Fig_6}
\end{figure}

\begin{figure}[tbp]
\includegraphics[width=85 mm]{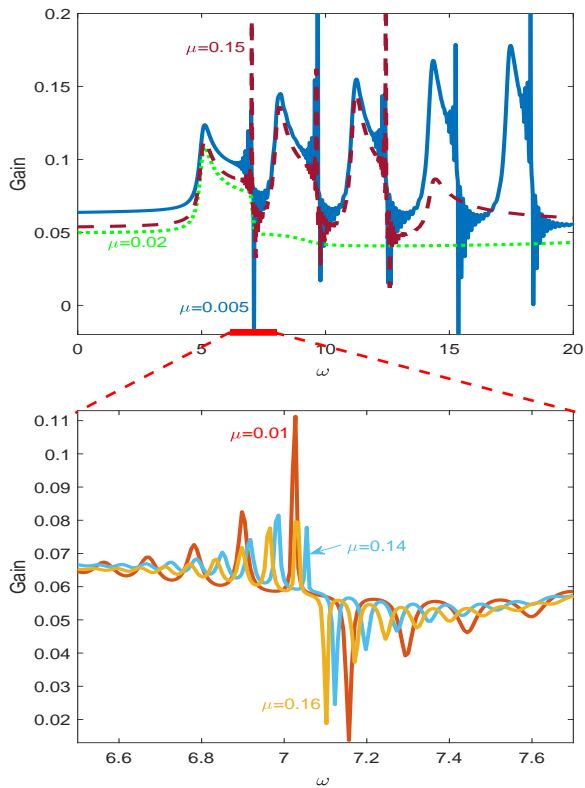} 
\caption{ {Color online. \textit{Top panel:} Gain as function of frequency $\protect%
\omega = 2\protect\pi f$ in GQD computed for the stripe width $W = 2 h v_{\mathrm{F}}/ \Delta$, temperature $T=1.3 \Delta /k_{\mathrm{B}}$, the ineleastic collision rate $\eta =0.15 \Delta$, the level spacing (Stark splitting) $2\Delta $ and the ac field
amplitude $E_{\mathrm{ac}}^{(0)} = 2.3 \Delta/e W$. The respective
values of electrochemical potential $\protect\mu$ inside the active region
are shown in figure. \textit{Bottom panel:} Blowup of the frequency region
denoted by red in the top panel that detalizing the fine structure of the
gain function near the lowest resonant frequency.}}
\label{Fig_7}
\end{figure}

\begin{figure}[tbp]
\includegraphics[width=85 mm]{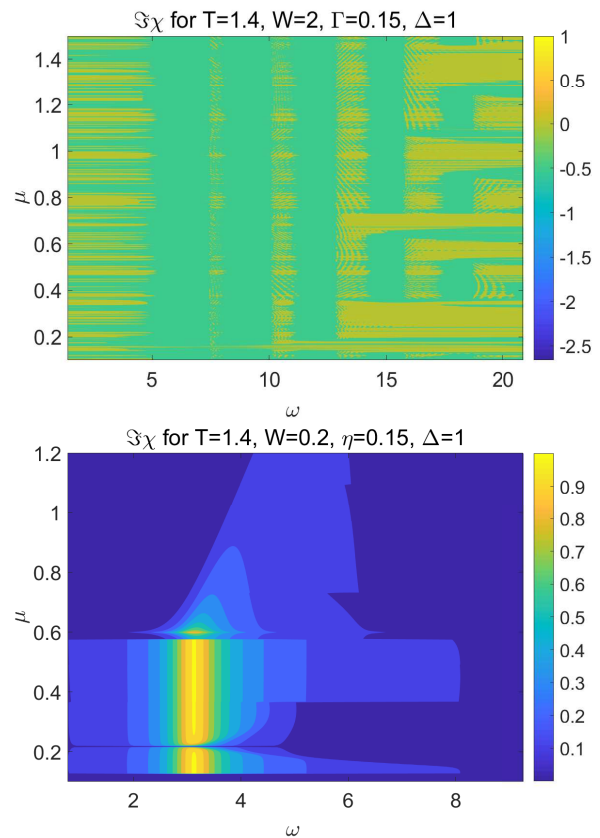} 
\caption{ {Color online. {\it Top panel:} Contour plot illustrating the sign switching of $%
\Im \protect\chi(\protect\mu, \protect\omega) $ for the graphene quantum dot
versus the frequency $\protect\omega $ and electrochemical potential $%
\protect\mu $ (both in units of $\Delta $) on the larger scale. Here the
green regions correspond to $\Im \protect\chi < 0$ while yellow areas
to $\Im \protect\chi > 0$. The GQD parameters are the same as in Fig.~7.
{\it Bottom panel:}  Contour plot $\Im \protect\chi(\protect\mu, \protect\omega) $ for much narrower stripe $W = 0.2 h v_{\mathrm{F}}/ \Delta$. The dependence is much smoother but the sign does not change.}}
\label{Fig_8}
\end{figure}

To further illustrate the capability of the graphene quantum dot to forming
the favorable lasing conditions we plot the dependenlies $\Re \{\chi (\mu
)\} $ and $\Im \{\chi (\mu )\}$ on the electrochemical potential $\mu $ in
Fig.~\ref{Fig_6}. From Fig.~\ref{Fig_6} one can see that by changing $\mu $, one alters the
shape of the susceptibility curves $\chi (\mu )$ considerably, thereby
enabling the flexible control over the coherent T-ray emission. Furthermore, the fast switchings of $\Re \{\chi (\mu )\} $ and $\Im \{\chi (\mu )\}$ takes place when $\mu $ hits the quantized level positions, which have narrow spacing and are dense for the relatively broad stripe $W = 2 h v_{\mathrm{F}}/ \Delta $.
The gain versus frequency is shown in Fig.~\ref{Fig_7} (top panel). We also
present more detailed plot in a narrower frequency region Fig.~\ref{Fig_7}
(bottom panel). From this Fig.~\ref{Fig_7} one can see that for the listed
GQD parameters (i.e., the stripe width $W = 2 h v_{\mathrm{F}}/ \Delta$, 
temperature $T=1.4 \Delta /k_{\mathrm{B}}$, the ineleastic collision rate $\eta =0.15 \Delta$ and the Stark splitting $2\Delta $), the gain exceeds the cavity loss. Provided $\Re \{\chi
(\omega ,\mu )\}\sim 0$, which is the case in certain intervals of $\mu $,
one achieves the necessary conditions for the coherent T-ray emission.

A general insight into the sign switching of the optical susceptibility for the graphene quantum dot versus $\omega $ and $\mu $ is given in the contour plot of $\Im \chi \left( \omega, \mu \right) $ as shown in Fig.~\ref{Fig_8} (top panel) where the green regions correspond to $\Im \chi < 0$ while yellow areas to $\Im \chi > 0$. Interesting, the sign switch does not happen for much narrower stripes with $W = 0.2 h v_{\mathrm{F}}/ \Delta$ where the level spacing is wide, although the whole dependence $\Im \chi \left( \omega, \mu \right) $ becomes much smoother as shown in Fig.~\ref{Fig_8} (bottom panel).

When designing the graphene THz lasers, there are other potential issues as
follows. (a) Pumping of non-equilibrium electron and hole excitations into
the active region of the THz laser causes not only inverse population of the
e/h-levels. An adverse side effect is that the non-equilibrium electrons and
holes eventually transfer their excessive energy to the lattice oscillations
and to other excitations in the system. This leads to an overall heating of
the active region during the induced emission process. Excessive Joule
heating of the active region might change its properties and can even
adversely impact the overall performance of the THz laser. Therefore, one
should pay attention to reducing the unwanted heating. The adverse heating
can be diminished by implementing the active region with an appropriate
geometry, crystallographic orientation, and dimensions. In this way one
eliminates certain electron-phonon scattering processes, e.g., by using a
stripe-shaped active region with zigzag edges. Similarly, one excludes the
indirect inter-level transitions which cause the acoustic phonon emission.
The remaining contribution originates solely from the direct inter-level
electron-hole recombination processes providing emission the THz photons out
the active region. (b) Forming an optimal energy spectrum inside the active
region. An increased width of the electron energy levels restricts the device
performance, widens overall frequency interval and causes line broadening of
the generated T-beam. One solution is to designing an active region with
narrow ($\eta <<\Delta $) quantized energy levels. It can be accomplished
by placing the appropriate split gates right on the top of graphene sheet
(see Fig.~\ref{Fig_2}a). An example of the energy diagram of the gate/open
graphene region boundary is sketched in Fig.~\ref{Fig_2}b). The electric
potential penetrates from the gate region into the open graphene on the
Debay screening length 0.5-2 nm, depending on the temperature and the charge
carrier concentration. Figs.~\ref{Fig_2}c,d show the electron subband
structure and the local electron density of states in the active region
respectively. (c) A distinguished feature of graphene is anisotropy of the
microscopic transport. Therefore one should design the active region with
appropriate dimensions and orientation. In this way, the major THz laser
parameters can be well defined during the fabrication process.

There are several reasons why the graphene THz lasers have a remarkable potential as compared to their conventional semiconducting counterparts. (\textit{i})~The intrinsic coherence in graphene is preserved
far better than in other non-superconducting electronic materials. It
happens due to so-called pseudospin conservation which is an intrinsic
feature of graphene. In particular, the good intrinsic coherence helps to
reducing the intrinsic noises. (\textit{ii})~The energy relaxation in
graphene is typically much slower than in other conductors. It allows
achieving a considerable degree the inverse level population. (\textit{iii})~Technically, the energy dependence of the election density of states in the 2D graphene enables manipulating of their properties by mere applying electric potentials to the gate electrodes. Furthermore, the 2D
geometry is well suitable to fabricating the bottom, top and side gate
electrodes. Owing to (\textit{i}) - (\textit{iii}), GQD comprises a system
with the robust, voltage-controlled narrow quantized energy levels, a
considerable inverse level population, good accumulation of the pumped
energy, which generates a very strong THz monochromatic beam.

The dissipative processes inside the active region cause fluctuations that
can be approximately described in terms of the quantum mechanical Langevin
equations. In this way one finds that the noise terms due to spontaneous
emission is $\propto r_{sp}$ and the nonradiative transition noise is $%
\propto r_{nr}$.

\section{Non-linear electromagnetic properties of GQD}

The unconventional excitation spectrum of the chiral fermions in the graphene stripe results in non-linear electromagnetic properties of this the two-dimensional atomic monolayer material. In particular, graphene has remarkable non-linear properties in the terahertz (THz) frequency range. These
create many suggestions for novel photonic devices, such as THz devices \cite%
{D-Sun}, optical modulators \cite{Liu}, photodetectors \cite{Mueller} and
polarizers \cite{Q-L-Bao}. One promising direction is exploiting the
non-linear electromagnetic response of the graphene stripe to an ac
electromagnetic field. Such the nonlinear effect might be used for the
frequency multiplication or for self-focusing of two dimensional Townes-like
solitons in the electrically tunable metastructure shown in Fig.1. Doping
graphene by applying the electric potentials to the gate electrodes allows
fine-tuning the nonlinear properties of such the metastructure. Total
internal reflection at the boundaries of the dielectric waveguide causes the
confinement of the $E$-field along the lateral $x$-direction. As the $y$%
-coordinate is varied along the stripe axis, the normalized $E$-field
cross-section along the $x$-direction changes. Such the change, which is
larger than in planar nonlinear waveguides corresponds to a significant
nonlinear optical current supported by the 2D graphene stripe. Below we find
that the third-order susceptibility in the graphene stripe is large enabling
to form the TE and TM spatial optical solitons. Stable Townes-like spatial
solitary waves propagating in the longitudinal direction originate from the
intraband current dominating the electron dynamics for THz excitations of
doped graphene. Significant magnitudes of the nonlinear optical
susceptibilities in the 2D graphene sheets were theoretically predicted in
Refs.~\cite{S-A-Mikhailov}, \cite{K-L-Ishikawa}. They have been
experimentally observed for third-order nonlinear effects by authors of Ref. 
\cite{E-Hendry}.

A nonlinear effect utilized to controlling light propagation at the micro-
and nano-scales is the formation of temporal and spatial EM-solitons \cite%
{R-W-Boyd}. We analyze the respective non-linear contribution for the
graphene strip in the geometry shown in Fig.~\ref{Fig_1}. Consider a
classical 2D particle with the charge $-e$ and the energy spectrum (\ref{Egr}%
), (\ref{kt}) as for a chiral fermion in the graphene stripe with zigzag
edges exposed to the time-dependent harmonic $y$-polarized electric field $%
E_{\mathrm{ac}}(t)=E^{0}_{\mathrm{ac}}\cos \Omega t$. The relevant
excitations are electrons in the vicinity of one gap edge while taking into
account the presence of two non-equivalent gap regions in the Brillouin zone
by introducing the valley-degeneracy factor $g_{v}=2$. According to the
Newton equation of motion 
\begin{equation}
\frac{dk_{y}}{dt}=-\frac{e}{\hbar }E_{\mathrm{ac}}(t)  \label{dkyt}
\end{equation}%
where we assume that the ac field is polarized along the stripe $y$-axis. In
Eq.~\ref{dkyt}, the momentum $k_{y}(t)$ is given by 
\begin{equation}
k_{y}(t)=k_{0}(t)=\varepsilon \sin \Omega t\text{,}  \label{kyt}
\end{equation}%
where $\varepsilon =eE^{0}_{\mathrm{ac}}/\hbar \Omega $.

In conventional 2D electron systems with the parabolic energy dispersion,
the velocity $v_{y}$, and hence, the current $j_{y}=-en_{s}v_{y}$ are
proportional to $\hbar k_{y}$, so that the normal 2D system responds at the
same frequency where $n_{s}$ is the areal density of change carriers. This
is different for the graphene stripe where the velocity
\begin{widetext}
\begin{equation}
v_{y}=\frac{1}{\hbar }\frac{\partial E_{p}}{\partial k_{y}} = v_{\mathrm{F}} 
\frac{k_{y}}{ \sqrt{k_{x}^{2}+k_{y}^{2}(t)+\left( \Delta /v_{\mathrm{F}} \right) ^{2}}}=v_{%
\mathrm{F}} \frac{\varepsilon \sin \Omega t}{\sqrt{k_{x}^{2} + \varepsilon^2
\sin^2{\Omega t}+\left( \Delta /v_{\mathrm{F}} \right) ^{2}}}  \label{vx1}
\end{equation}%
\end{widetext}
is not merely proportional to $k_{y}$. In the extreme limit, when $k_{x}$
and $\Delta /v_{\mathrm{F}}$ in Eq.~(\ref{vx1}) are close to zero, $v_{y}$ is
proportional to $\mathrm{sgn}(p_{x})$ and the ac electric current $%
j_{y}=-en_{s}v_{y}$ has anharmonic contributions 
\begin{equation}
j_{y}\left( t\right) =en_{s} v_{\mathrm{F}}\frac{4}{\pi }\left\{ \sin \Omega
t+\frac{1}{3}\sin 3\Omega t+\frac{1}{5}\sin 5\Omega t+...\right\} \text{.}
\label{jx}
\end{equation}

Both, the gate voltage and chemical doping can shift the chemical potential $%
\mu $ of electrons in graphene to the upper $E_{p2}$ or to the lower $E_{p1}$
band. Let us assume that the chemical potential $\mu $ lies in the upper
band $E_{p2}= v_{\mathrm{F}} p$, the temperature is small, $k_{B}T<<\mu $,
and the system is subjected to the time-dependent ac electric field $E_{%
\mathrm{ac}}(t)$. Then the momentum distribution function of electrons $%
f_{p}(t)$ is described \cite{S-A-Mikhailov} by Boltzmann equation%
\begin{equation}
\frac{\partial f_{p}(t)}{\partial t}-\frac{\partial f_{p}(t)}{\partial 
\mathbf{p}}e\mathbf{E}_{\mathrm{ac}}(t)=0\text{,}  \label{BE2}
\end{equation}%
where we have disregarded collisions of electrons with impurities, phonons
and other lattice imperfections. Equation (\ref{BE2}) has the exact solution%
\begin{equation}
f_{p}(t)=\mathcal{F}_{0}\left( \mathbf{{p}-{p}_{0}\left( t\right) }\right) 
\text{,}
\end{equation}%
where%
\begin{equation}
\mathcal{F}_{0}\left( \mathbf{p}\right) =\frac{1}{\exp \left( \frac{ v_{%
\mathrm{F}}p-\mu }{T}\right) +1}
\end{equation}%
is the Fermi-Dirac function, and%
\begin{equation}
\mathbf{{p}_{0}\left( t\right) =}-e\int_{-\infty }^{t}dt^{\prime }\mathbf{E}%
_{\mathrm{ac}}(t^{\prime })
\end{equation}%
is the solution of the single particle classical equation of motion. Thus,
the former equations derived in previous sections remain valid provided we
replace $\mathbf{{p}\rightarrow {p}-{p}_{0}}\left( t\right) $ in the
respective distribution functions.

The non-linear regime (\ref{jx}) is achieved at $|p_{0}|\gg p_{F}$, or at%
\begin{equation}
\mathcal{E}\equiv \frac{eE_{\mathrm{ac}}^0 v_{\mathrm{F}}}{\Omega \mu }>>1%
\text{.}  \label{E-cal}
\end{equation}%
According to Eq.~(\ref{E-cal}), the non-linear effect becomes essential
alredy at $E_{\mathrm{ac}}^0 \geq $ 1.4 kV/cm provided $f=\Omega/2\pi = 1$%
~THz and $\mu = 0.06$~eV. Meaning of the above relationship (\ref{E-cal}) is
that the energy, gained by electrons from the ac field during the
oscillation period should be large as compared to their average equilibrium
energy. In the low-field limit, the response is linear (i.e., the $j(t)$ dependence
has a sinusoidal form), while at strong fields the time dependence of the
current tends to that given by Eq. (\ref{jx}). The strong-field condition (%
\ref{E-cal}) can be rewritten as%
\begin{equation}
E_{\mathrm{ac}}^0>>\frac{2 \Omega \sqrt{\pi n_{s}}}{e\sqrt{g_{s}g_{v}}}
\label{str-fld}
\end{equation}%
which means that the required ac electric field grows linearly with the
electromagnetic wave frequency and with the square root of the electron
density.

There are following limitations on applicability of the quasi-classical
method to describing the electromagnetic response of graphene stripe.
Physically, using the Newton equation (\ref{dkyt}), one takes into account
contribution the intra-band transitions to the ac electric current while
ignoring the inter-band transitions between the lower quasi-hole and the
upper quasi-electron bands. This is only possible if the frequency of the
electromagnetic radiation satisfies the inequality
\begin{equation}
\Omega \ll \mathrm{max}\{\mu ,T\}\text{.}
\end{equation}%
At room temperature and for the electric charge carrier densies $n_{s}\simeq
10^{11}-10^{12}$ cm$^{-2}$ the above inequality limits the frequency band to 
$\sim 10-30$ THz.

We estimate of the third-order susceptibility $\chi _{\mathrm{gr}}^{(3)}$ by
computing the relevant Fourier coefficients of the time-dependent $\chi
\left( t ,\mathbf{E}_{ext}\left( t\right) \right)$ as 
\begin{equation}
\chi_{\mathrm{gr}}^{(n)} \left( \Omega ,\mathbf{E}_{ext}^0 \right) =\frac{%
\Omega}{2 \pi} \int_{0}^{2\pi /\Omega} \chi \left( t^{\prime } ,\mathbf{E}%
_{ext}\left(t^{\prime } \right) \right) e^{-i\Omega n t^{\prime }}
dt^{\prime } \text{,}  \label{harm}
\end{equation}%
where $\Omega $ is the ac field frequency, $n$ is the integer number. Likewise, we also expand the
electric current density in graphene stripe $j_{\mathrm{gr}}$ in powers of $%
\psi =eA/p_{F}$ (where A is the vector potential) up to the third order: 
\begin{equation}
j_{\mathrm{gr}}\simeq \frac{\psi }{\sqrt{1+\psi ^{2}}}\simeq \psi -\frac{%
\psi ^{3}}{8}\text{,}
\end{equation}%
finding the nonlinear third order intraband current density $j_{\mathrm{gr}%
}^{(3)}$.

\begin{figure}[tbp]
\includegraphics[width=85 mm]{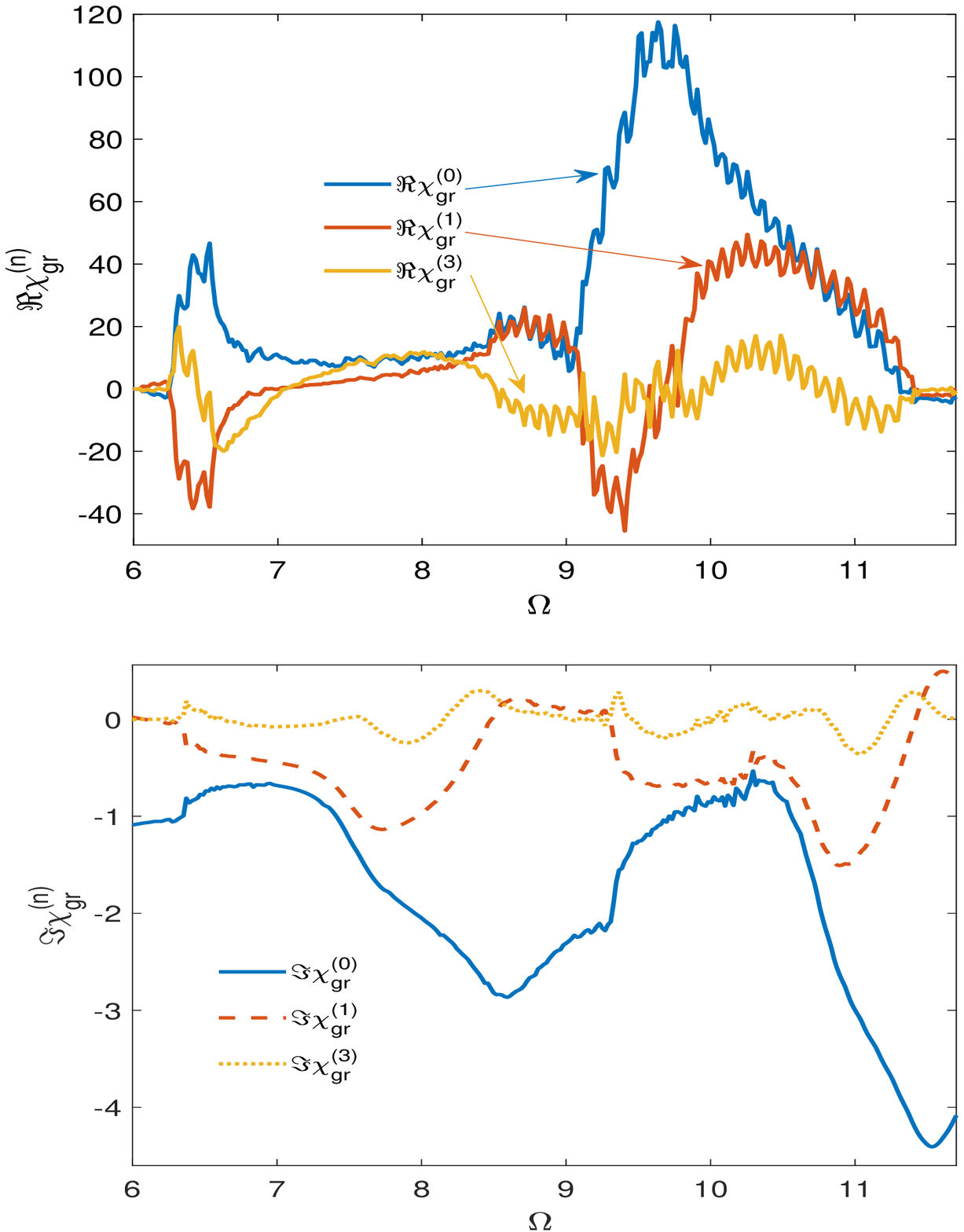} 
\caption{ {Color online. \textit{Top panel:} The $\Re \protect\chi_{\mathrm{%
gr}}^{(n)} \left( \Omega ,\mathbf{E}_{ext}^0 \right) $ for the harmonics $%
n=0,1,3$ as functions of the ac field frequency $\Omega $. The GQD
parameters are as follows: The stripe width $W = 2 h v_{\mathrm{F}}/ \Delta$, temperature $T=1.2 \Delta /k_{\mathrm{B}}$, the ineleastic collision rate $\eta =0.12 \Delta$, the level
spacing (Stark splitting) $2\Delta $ and the ac field
amplitude $E_{\mathrm{ac}}^{(0)} = 2.3 \Delta/e W$. \textit{%
Bottom panel:} Respective plots for $\Im \protect\chi_{\mathrm{gr}}^{(n)}
\left( \Omega ,\mathbf{E}_{ext}^0 \right) $. One can see that the magnitudes
of all the harmonics are comparable with each other.}}
\label{Fig_9}
\end{figure}

Harmonics of the ac field-dependent non-linear susceptibility $\chi_{\mathrm{%
gr}}^{(n)} \left(\Omega ,\mathbf{E}_{ext}^0 \right) $ are computed
numerically using the above formulas. In Fig.~\ref{Fig_9} we show the
Fourier components of the current versus the field parameter $\mathcal{E}$ given by Eq.~(\ref{E-cal}).
When $\mathcal{E}$ becomes larger than $\simeq 4$, the Fourier amplitudes
saturate and one gets in the ultimate non-linear regime. From the plots of
susceptibility harmonics $\Re \chi_{\mathrm{gr}}^{(0,1,3)} \left( \Omega ,%
\mathbf{E}_{ext}^0 \right) $ and $\Im \chi_{\mathrm{gr}}^{(0,1,3)} \left(
\Omega ,\mathbf{E}_{ext}^0 \right)$ illustrating the non-linear effects in
graphene stripe with zigzag edges shown in Fig.~\ref{Fig_9} one can see that
the 0-th, 1-st and 3-harmonics of $\chi \left( t^{\prime } ,\mathbf{E}%
_{ext}\left(t^{\prime } \right) \right) $ are about the same order of
magnitude, which suggests the significance of non-linear phenomenon in the GQD
system under consideration.

In conclusion of this Section, due to the unconventional dispersion law (\ref%
{Egr}), (\ref{kt}) of the chiral fermions, the response of graphene stripe
to an ac electromagnetic field is intrinsically non-linear.

\section{Time evolution of susceptibility}
An accurate estimation of the time required to reach the steady state regime
represents a tedious task involving a self-consistent solution of a complex
system of the Boltzmann equations complemented by equations for the HF
excitation spectrum. Besides, the equations must be complemented by respective
boundary conditions defining the geometry and initial state. In this work we
provide just a simplest insight how the optical susceptibility $\chi
^{\prime \prime }\left( \omega ,t\right) $ evolves in response to a pulse of
injection current incurring a sharp change of the effective electron
temperature $T^{\ast }$ of GQD. Let us assume that the pulse of injection
current heats GQD, whose effective temperature increases up to $T^{\ast }$. Basically, after the
pulse ends, the time evolution of the HF distribution function $f(\epsilon ,t)$ is found as a
solution of Boltzmann equation. For the sake of simplicity, we use the
effective temperature approximation that gives%
\begin{equation}
f(\epsilon ,t)=\frac{1}{\exp \left( \frac{\epsilon -\mu }{T^{\ast }\left(
t\right) }\right) +1} {\text ,}  \label{fEt}
\end{equation}%
where the time dependence of $T^{\ast }$ is determined by the energy relaxation of the HF distribution function. In the above approximation we use%
\begin{equation}
T^{\ast }\left( t\right) = T \exp \left[ \left( -\frac{t-t_{0}}{\tau _{p}}\right)
^{n}\right]  {\text ,}   \label{jt}
\end{equation}%
where we take $n = 100$ and $t_{0}=\tau _{p}=2.5$~ns.  
\begin{figure}[tbp]
\includegraphics[width=85 mm]{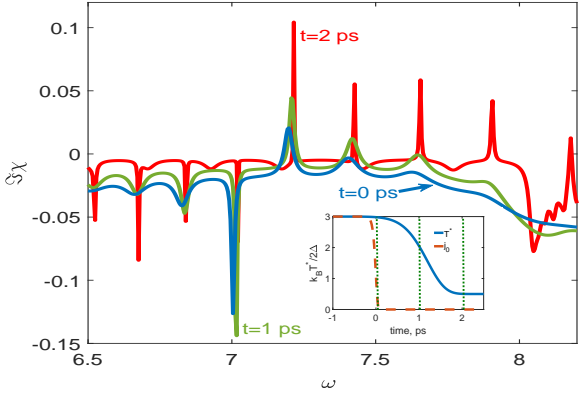} 
\caption{ {Color online. The imaginary part of optical susceptibility $\Im \chi \left( \omega ,t \right) $ as functions of the frequency $\omega $ at time moments $t =$0~ps, 1~ps and 2~ps after the injection current pulse ends. Here we used the GQD stripe width $W = 2 h  v_{\mathrm{F}}/ \Delta$, the steady state temperature $T=0.5 \Delta /k_{\mathrm{B}}$, the level spacing (Stark splitting) $2\Delta $ and  the energy relaxation time $\tau_{\rm HF}=$1~ps. One can see that the steady state of GQD is achieved at $t \sim 2$~ps, after the injection current pulse ends and the HF excitations recombine in the energy. Inset shows the time dependence of the effective temperature $T^*(t)$ (solid blue line) that achieves its steady state value $T = 0.5\Delta/k_B $ at $t \sim 2$~ps, after the injection current pulse (red dash) ends at $t=0$.}}
\label{Fig_10}
\end{figure}
When the energy recombination time $\tau_{\rm HF} $ of the HF excitations is very short, $\tau_{\rm HF} << {\rm min}\{\tau _{p}, t_{0}\}$ (typically $\tau_{\rm HF} \sim 10^{-12}$~s), the time evolution of $T^{\ast }\left( t\right) $ immediately follows the change of injection current, $T^{\ast }\left( t\right) \propto j\left( t\right) $. However, the scenario becomes different when $\tau_{\rm HF} \sim \{\tau _{p}, t_{0}\}$. In the last case, when the injection current pulses are sufficiently short, the optical susceptibility reaches the steady state on the timescale $ \sim \tau_{\rm HF}$. Such the time evolution of $\Im \chi \left( \omega ,t \right) $ is illustrated in Fig.~\ref{Fig_10} by using the above simple model (\ref{fEt}) and (\ref{jt}) allowing to determine the behavior of the graphene \textquotedblleft particle\textquotedblright\ from the time that the injection current pulse ends ($t = $0~ps) till reaching steady state at $t = $2~ps. The graph in Fig.~\ref{Fig_10} helps to understand and evaluate the role of non-equlibrium effect in GQD.

\section{Conclusions}

Obtained results suggest that the considered design of the graphene quantum dot (GQD) allows the all-electrical control of the optical susceptibility $\chi (\omega ,\mu )$. This becomes possible because the magnitude and sign of both the real and imaginary parts of $\chi (\omega ,\mu )$ depend on the electrochemical potential $\mu $ and on the frequency variable $\omega $. Technically, in the GQD structure one can change the magnitudes of $\Delta $ and $\mu $ by applying appropriate electric potentials to the local gates as depicted in Fig.~\ref{Fig_1}, thereby enabling the flexible control of the GQD optical properties.

Furthermore, likewise the pristine graphene, the GQD structure has
remarkable non-linear electromagnetic properties stemming from the
unconventional dispersion law of the chiral fermions in the graphene stripe
with zigzag edges. A strong non-linear effect arises because the optical
susceptibility depends on the ac field intensity. The magnitude of
high-order harmonics is significant even in relatively weak ac fields,
causing appearance of variety the non-linear effects.

In experiments, the THz radiation is detected in several ways. For instance,
one can form a Josephson junction in an adjacent area on the same substrate,
which will serve as a THz detector. Another option is to deposit GQD THz
detector on the same substrate next to the GQD THz emitter. Furthermore, one
can use metallic co-planar strip lines as THz antennas to detect the T-rays.
Special attention must be paid to creating of the sharp and narrow electron
quantized levels formed between two timber-like multilayered gate electrodes
deposited along the ZZ-direction. Such the energy levels are robust in
presence of lattice defects and imperfections remaining to be very narrow
and sharp since they are topologically protected \cite%
{TEbook,Arabs,Shafr-Graph-Book}. In the active region of the THz emitter, the
coherent monochromatic THz waves originate from the quantum
transitions between the sharp localized levels. The energy level splitting is
readily controlled by the voltage difference between the gate electrodes
having a multilayered structure as depicted in Fig.~\ref{Fig_1}.

The unconventional electromagnetic (EM) properties of the graphene quantum dots (GQD) have a promising potential for practical applications. Experimentally, it would be interesting to fabricating the all-electrically controlled GQD based on the graphene stripes with atomic zigzag edges. The calculation results of the optical susceptibility indicate strong dependence on the frequency and electrochemical potential, which can be exploited for experimental observing the tunable THz emission. Another result is the non-linear electromagnetic response of GQD, whose mechanism is related to the unconventional properties of chiral fermions in graphene stripes with atomic zigzag edges. The obtained data allow better understanding the physical mechanisms related to the electrically controlled GQD showing remarkable electromagnetic properties.

\section{Appendix}
\subsection{Susceptibility of pristine graphene}
Here we provide technical details concerning the optical susceptibilty of pristine graphene.
We use  Eq.~(\ref{kappa1}), which is written for indirect interband transitions accompanied by
the absorption/emission of a phonon with the finite momentum $\mathbf{{q}%
\neq 0}$. Here we use units with $\hbar =1$ and $k_{\mathrm{B}}=1$
if not stated otherwise. When the absorption/emission process is accompanied
by the photon instead of the phonon, the interband transitions become direct
and the above Eq.~(\ref{kappa1}) is simplified. For the photons, whose
absorption creates the electron-like HF (below for brevity we call them
electrons) in the conductive band while the hole-like HF (we call them
holes) in the valence band (direct interband transitions), we set $\mathbf{{q%
}=0}$. Then, using that $ss^{\prime }=-1$ we get the respective energy
change as $\epsilon _{s^{\prime },v}-\epsilon _{s,c}=\pm 2\epsilon $ and
from Eq. (\ref{kappa1}) we get 
\begin{widetext}
\begin{eqnarray}
\chi \left( \omega \right) &=&-\frac{6\sqrt{3}\left\vert d_{cv}\right\vert
^{2}}{\pi v_{F}^{2}}\int_{0}^{v_{\mathrm{F}}\Lambda }d\epsilon \epsilon
\left( f\left( -\epsilon \right) -f\left( \epsilon \right) \right)
\label{kappa3} \\
&&\times \left( \frac{1}{\omega +2\epsilon +i\eta }-\frac{1}{\omega
-2\epsilon +i\eta }\right) \text{,}
\end{eqnarray}%
where $\Lambda \approx 1/a$ is the cutoff momentum and $a$ is the lattice
constant.

In beginning we find how the steady state functions $\chi \left( \omega
\right) $ and $\chi \left( \mu \right) $ depend respectively on the photon
frequency $\omega $ and the HF electrochemical potential $\mu $. In the
above formulas we set 
\begin{equation}
f\left( \epsilon \right) =\frac{1}{e^{\left( \epsilon -\mu \right) /T}+1}%
\text{.}
\end{equation}%
In the zero--temperature limit $T=0$ we simply set $f\left( \epsilon \right)
=\theta \left( \mu -\epsilon \right) $ and $f\left( -\epsilon \right) =1$
for electron doping with $\mu >0$, where $\mu $ is the electron
electrochemical potential. Respectively, for the hole doping we get $\mu <0$%
. Then%
\begin{equation}
\chi \left( \omega \right) =-\frac{6\sqrt{3}\left\vert d_{cv}\right\vert ^{2}%
}{\pi v_{F}^{2}}\int_{\mu }^{v_{F}\Lambda }\left( \frac{1}{\omega +2\epsilon
+i\eta }-\frac{1}{\omega -2\epsilon +i\eta }\right) \epsilon d\epsilon \text{%
,}  \label{kappa4}
\end{equation}%
where $\zeta ^{2}=\left( \eta ^{2}-2i\eta \omega -\omega ^{2}\right) /4$ and
we have used%
\begin{equation}
f\left( -\epsilon \right) -f\left( \epsilon \right) =1-\theta \left( \mu
-\epsilon \right) =\left\{ 
\begin{array}{c}
0\text{ for }\epsilon <\mu \\ 
1\text{ for }\epsilon >\mu%
\end{array}%
\right\vert \text{.}
\end{equation}%
An immediate integration of Eq. (\ref{kappa4}) gives a simple analytical
expression in the form 
\begin{equation}
\chi \left( \omega ,\mu \right) =-\frac{i}{\pi }\frac{3\sqrt{3}\left\vert
d_{cv}\right\vert ^{2}}{2\pi v_{F}^{2}}\left[ \Lambda -\mu +\frac{\omega
+i\eta }{2}\left( \tanh ^{-1}\left( \frac{2\mu }{\omega +i\eta }\right)
-\tanh ^{-1}\left( \frac{2\Lambda }{\omega +i\eta }\right) \right) \right] 
\text{.}  \label{kappa5}
\end{equation}%

The above calculations are illustrated in Figs. \ref{Fig_3} where we
show the real $\chi ^{\prime }=\Re \chi $ and imaginary $\chi ^{\prime
\prime }=\Im \chi $ parts of the optical susceptibility versus the frequency 
$\omega $ and electrochemical potential $\mu $. 
\begin{equation}
\chi \left( \omega ,\mu \right) =-\frac{6\sqrt{3}\left\vert
d_{cv}\right\vert ^{2}}{\pi v_{F}^{2}}\left[ v_{F}\Lambda -\mu +\frac{\omega 
}{4}\left( \ln \left\vert \frac{\omega +2\mu }{\omega -2\mu }\right\vert
+i\pi \theta \left( \omega -2\mu \right) \right) \right] \text{.}
\end{equation}%
It is also instructive to find the number of excitations $n\left( \mu
\right) $ in the 2D graphene. For the equilibrium case we find 
\[
n=\frac{N}{L^{2}}=-\frac{12\sqrt{3}}{\pi v_{F}^{2}}\int_{0}^{v_{F}\Lambda
}\epsilon f\left( \epsilon \right) d\epsilon =-\frac{12\sqrt{3}}{\pi
v_{F}^{2}}\int_{0}^{v_{F}\Lambda }\epsilon \frac{1}{e^{\beta \left( \epsilon
-\mu \right) }+1}d\epsilon -\frac{12\sqrt{3}}{\pi v_{F}^{2}\beta ^{2}}%
\int_{0}^{\beta v_{F}\Lambda }\epsilon \frac{1}{e^{\left( \epsilon -\beta
\mu \right) }+1}d\epsilon \text{.} 
\]%
Then one gets 
\begin{equation}
n\left( \mu \right) =-\left( 12\sqrt{3}T^{2}/\pi v_{F}^{2}\right)
[Li_{2}\left( -e^{-\frac{\mu }{T}}\right) -Li_{2}\left( -e^{\frac{\Lambda
-\mu }{T}}\right) +\Lambda \left( \Lambda -2T\log \left( e^{\frac{\Lambda
-\mu }{T}}+1\right) \right) /T^{2}]\text{.}
\end{equation}
\end{widetext}
In more realistic conditions, e.g., when the temperature is finite while the graphene sheet is deposited on a substrate and its electronic states are controlled by gate electrodes, one finds the optical susceptibility numerically, as described in the main text. Likewise, numeric solutions also used for studying the non-stationary and non-equilibrium properties of the graphene samples.

\end{document}